\documentclass[aps,pra,twocolumn,superscriptaddress,showpacs,reprint]{revtex4-1}
\usepackage{graphicx}
\usepackage{epstopdf}
\usepackage{amsfonts}
\usepackage{amssymb}
\usepackage{amsmath}
\usepackage{hyperref}
\usepackage{color}
\usepackage{braket}
\usepackage{soul}
\newcommand{\pjct}[2]{\left\vert#1\right\rangle\!\left\langle#2\right\vert}
\newcommand{\tn}[1]{\textnormal{#1}}
\newcommand{\abs}[1]{\left|#1\right|}

\def\Tr{\mbox{Tr}}

\begin{document}

\title{
Quantum macroscopicity measure for arbitrary spin systems and its application to quantum phase transitions
}

\author{Chae-Yeun Park}
\affiliation{Center for Macroscopic Quantum Control, 
Department of Physics and Astronomy,
Seoul National University, Seoul, 08826, Korea}  

\author{Minsu Kang}
\affiliation{Center for Macroscopic Quantum Control, 
Department of Physics and Astronomy,
Seoul National University, Seoul, 08826, Korea}  

\author{Chang-Woo Lee}
\affiliation{Center for Macroscopic Quantum Control, 
Department of Physics and Astronomy,
Seoul National University, Seoul, 08826, Korea}  
\affiliation{School of Computational Sciences, Korea Institute for Advanced Study, Seoul 02445, Korea}

\author{Jeongho Bang}
\affiliation{Center for Macroscopic Quantum Control, 
Department of Physics and Astronomy,
Seoul National University, Seoul, 08826, Korea}  

\author{Seung-Woo Lee}
\affiliation{Center for Macroscopic Quantum Control, 
Department of Physics and Astronomy,
Seoul National University, Seoul, 08826, Korea}  
\affiliation{Quantum Universe Center, Korea Institute for Advanced Study, Seoul 02445, Korea}

\author{Hyunseok Jeong}
\email{h.jeong37@gmail.com}
\affiliation{Center for Macroscopic Quantum Control, 
Department of Physics and Astronomy,
Seoul National University, Seoul, 08826, Korea}

\date{\today}

\begin{abstract}
We explore a previously unknown connection between two important problems in physics, i.e., quantum macroscopicity and the quantum phase transition.
We  devise a general and computable measure of quantum macroscopicity that can be applied to arbitrary spin states.
We find that a macroscopic quantum superposition of an extremely large size arises during the quantum phase transition of the transverse Ising model
in contrast to some seeming macroscopic quantum phenomena such as superconductivity, superfluidity and Bose-Einstein condensates. 
Our result may be an important step forward in understanding macroscopic quantum properties of many-body systems.
\end{abstract}

\maketitle

%%%%%%%%%% INTRODUCTION %%%%%%%%%%
%\textit{Introduction.}--
\section{Introduction}
Quantum mechanics does not preclude the possibility of a macroscopic object being in a quantum superposition \cite{Schrodinger1935}.
There has been interesting progress
in generating macroscopic superpositions using 
atomic and molecular systems \cite{MonroeCat,C60},
superconducting circuits \cite{SQUID1,SQUID2}, and
optical setups \cite{Ourjoumtsev,Gao,Afek}.
On the other hand, how to sharply define and quantify ``quantum macroscopicity'' is 
a slippery issue \cite{Leggett2002}.
As originally pointed out by Leggett \cite{Leggett1980}, a genuine macroscopic quantum superposition should be distinguished from a product state of many microscopic quantum superpositions.
In this sense, none of Debye's $T^3$ law, Bose-Einstein condensates, superconductivity and superfluidity is a genuine macroscopic quantum superposition
 \cite{Leggett1980,MacroReview}.

 There have been a number of studies on quantum macroscopicity of various physical systems
  \cite{Leggett1980,Leggett2002,Dur,Shimizu2002, Bjork,Shimizu2005, Korsbakken, Morimae2010, Mar,Cavalcanti, LeeJeong2011, Frowis2012NJP,Nimmrichter2013, Sekatski, FrowisLinking,MacroReview,arndt14, %Baumgratz2014, Girolami2014,
 Laghaout2014, Sekatski2014, SqueezedMacro}
   including multi-component and mixed states \cite{LeeJeong2011, Frowis2012NJP,Nimmrichter2013,FrowisLinking,MacroReview,arndt14}.  
    While most of the proposals for   quantification of quantum macroscopicity  are limited to specific forms of states, 
recently, more general measures were suggested for arbitrary bosonic systems based on interference fringes in phase space \cite{LeeJeong2011} and for arbitrary spin systems
based on the quantum Fisher information (QFI) \cite{Frowis2012NJP}.
However, it is  difficult to use the QFI based measure for large spin systems because of its computational complexity due to the requirement
of density matrix diagonalization \cite{Frowis2012NJP}.

Our study in this paper suggests that the quantum phase transition (QPT)
described in the transverse Ising model is a genuine macroscopic quantum phenomenon in contrast to other seeming macroscopic quantum phenomena such as superconductivity, superfluidity and Bose-Einstein condensates.
 In order to investigate genuine quantum macroscopicity of many-body quantum systems,
 we devise a general and computable measure that can be readily applied to arbitrary spin systems.
It is  based on interference fringes in phase space so that it has a strong conceptual connection with the one for bosonic systems \cite{LeeJeong2011}.
It has a similar mathematical structure as the QFI based measure \cite{Frowis2012NJP} and the two measures become identical for pure states.
However,  our measure
 requires less computational complexities than the QFI based measure because the density-matrix diagonalization is not needed (see Appendix~\ref{sec:time_compare}).
In search of genuine macroscopic quantum phenomena, we investigate many-body spin states undergoing the QPT  that is a classic many-body phenomenon.
The QPT of the transverse Ising model turns out to be a genuine macroscopic quantum phenomenon, and it suggests that a macroscopic quantum superposition of an extremely large size may arise during the QPT. 

\section{Quantum macroscopicity for arbitrary spin systems.}
It was shown that quantum macroscopicity of arbitrary harmonic oscillator states can be quantified based on its Wigner function structure in the phase space \cite{LeeJeong2011}. It is a separate problem to find whether this type of approach is possible for spin systems.
The Wigner (or Stratonovich-–Weyl) distribution for a spin-$S$ particle is represented by \cite{GroupTextbook}
\begin{align}
W({\bf n}) =\sqrt{\frac{4\pi}{2S+1}}\sum_{L=0}^{2S}\sum_{M=-L}^{L}  \ \chi_{L,M}^{(S)}  Y_{L,M}({\bf n}),
\label{eq:spin-Wig}
\end{align}
where $Y_{L,M}({\bf n})$ denotes spherical harmonics with a three dimensional unit vector ${\bf n}$=(sin$\theta$cos$\phi$, sin$\theta$sin$\phi$, cos$\theta$) and $\chi_{L,M}^{(S)}=\Tr[\hat{T}_{L,M}^{(S)\dagger}\rho ]$ is the characteristic function with the irreducible tensor operator $\hat{T}_{L,M}^{(S)}$.
The matrix elements of the  irreducible tensor operator are defined as
$ \bra{S,m'}\hat{T}_{L,M}^{(S)} \ket{S,m} =  \sqrt{(2L+1)/(2S+1)} C_{S,m; L,M}^{S,m'}$ with Clebsch-Gordan coefficients $C_{S,m; L,M}^{S,m'}$ and
an eigenstate  $\ket{S,m}$ of the $z$-component spin operator $\hat{S}_z$ with its eigenvalue $m$.
Here we work with a natural unit of $\hbar=1$.

In order to define a measure of quantum macroscopicity  for an arbitrary spin system, we attempt to simultaneously quantify both (i) the distinctness between the component states of a superposition state and (ii) the degree of genuine quantum coherence  between those component states against their classical mixture. 
An important observation is that  these properties are closely related to two quantities: the \textit{frequency} and the \textit{magnitude} of the interference fringes of the Wigner function. Remarkably, this is consistent with the case of harmonic oscillator systems that leads to the definition of quantum macroscopicity for bosonic states previously investigated in Ref.~\cite{LeeJeong2011}.

\begin{figure}[t]
\includegraphics[width=240px]{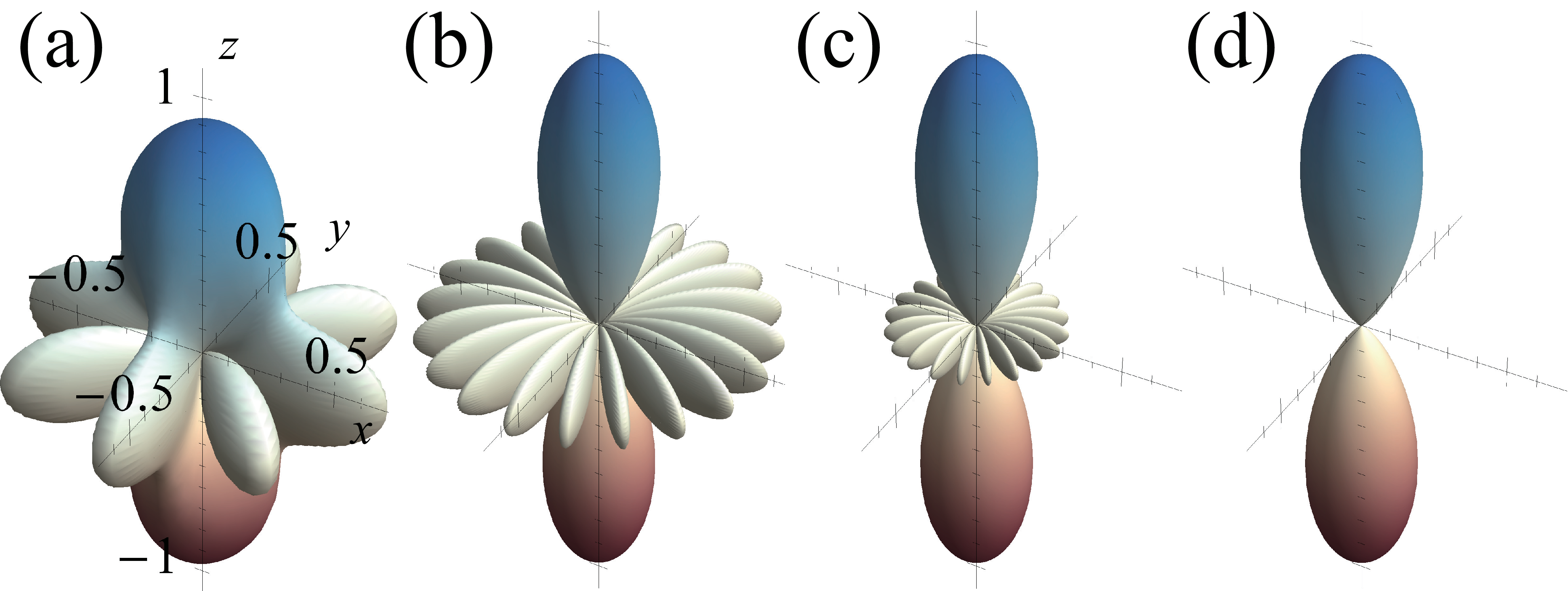}
\caption{(Color online)  Wigner distributions of $(\ket{S,S}+\ket{S,-S})/\sqrt{2}$ for (a) $S=2$ and (b) $S=5$. Each distribution consists of two peaks along the $\pm z$ direction and interference fringes around the $z$ axis. When the pure superposition with $S=5$ becomes a mixed state, $\{\pjct{5,5}{5,5}+\pjct{5,-5}{5,-5} + \gamma(\pjct{5,5}{5,-5} +\pjct{5,-5}{5,5})\}/2$,   the interference fringes are obviously reduced for (c)  $\gamma=1/2$ and they  completely disappear for  (d) $\gamma=0$.}
\label{fig:spin-Wig}
\end{figure}

To clarify, let us first consider an example of a superposition of two spin-$S$ components of opposite signs, $(\ket{S,S}+\ket{S,-S})/\sqrt{2}$.
We plot its Wigner distribution in a three-dimensional phase space for $S=2$ in Fig.~\ref{fig:spin-Wig}(a)  and for $S=5$ in Fig.~\ref{fig:spin-Wig}(b).
Once the Wigner distribution is given, the expectation value of an arbitrary spin operator $\hat{J}$ can be calculated by the overlap relation
$\langle \hat{J} \rangle = (2S+1)/(4\pi) \int d\Omega  W_J({\bf n}) W({\bf n}) $,
where $d\Omega=\sin{\theta}d\theta d\phi$ and $W_J({\bf n})$ is the Wigner representation of operator $\hat{J}$ defined in the same manner as the Wigner distribution by replacing the density operator $\rho$ by $\hat{J}$. 
As shown in Figs.~\ref{fig:spin-Wig}(a) and \ref{fig:spin-Wig}(b), the superposition of a larger value of $S$ shows more frequent periodic patterns around the $z$ axis.
These are the interference fringes of the superposition between the two component states, $\ket{S,S}$ and $\ket{S,-S}$, with  perfect quantum coherence.
When it undergoes dephasing, it becomes a mixed state as 
$(\ket{S,S}\bra{S,S}+\gamma\ket{S,S}\bra{S,\text{\small}{-}S}+\gamma\ket{S,\text{\small}{-}S}\bra{S,S}+\ket{S,-S}\bra{S,-S})/2$ with $0\leq\gamma<1$.
The magnitude of the fringes then becomes smaller as shown in Fig.~\ref{fig:spin-Wig}(c)
and it completely disappears when the state is fully decohered (i.e. $\gamma=0$) as in Fig.~\ref{fig:spin-Wig}(d).

The spherical harmonics $Y_{L,M}({\bf n})$ in Eq.~(\ref{eq:spin-Wig}) form a complete orthonormal basis to describe an arbitrary spin system.
The periodic interference fringes in Fig.~\ref{fig:spin-Wig}
are attributed to these spherical harmonics 
and the $\phi$ dependence of $Y_{L,M}({\bf n})$ appears only in the form of $\exp[{ i M\phi}]$.
It means that $M$ is the frequency of the interference fringes due to $Y_{L,M}({\bf n})$.
We then notice from Eq.~(\ref{eq:spin-Wig}) that the complex amplitude of a certain frequency component for $M$, i.e. $Y_{L,M}({\bf n})$, is  $\chi_{L,M}^{(S)}$.
Since the two essential elements of quantum macroscopicity  are identified, we can  define its measure to be proportional to $\sum (\tn{frequency})\times(\tn{magnitude})$.
Considering the normalization factor, we can straightforwardly attempt a simple definition
as
\begin{align}
\begin{split}
{\cal I}_z&=\frac{1}{2 S} \frac{1}{\mathcal{P}}\sum_{L=0}^{2S} \sum_{M=-L}^{L} M^2 \abs{\chi_{ L, M}^{(S)} }^2\\
&=
\frac{1}{2 S \mathcal{ P}}
\frac{2S+1}{4\pi }\int d\Omega \ W({\bf n})\hat{L}_z^2 W({\bf n}).
\label{eq:i-guess-wig}
\end{split}
\end{align}
where $\mathcal{P}(\rho)=
\sum_{L',M'}|\chi_{L',M'}^{(S)}|^2=\tn{Tr}[\rho^2]$ corresponds to the purity of the quantum state
and $\hat{L}_z=i\partial /\partial\phi$ (see the Appendix for the proof).
The scaling factor $(2S)^{-1}$ is introduced  to make ${\cal I}_z=S$ for $\gamma=1$.

However, the preliminary definition in Eq.~(\ref{eq:i-guess-wig}) is appropriate only for a single spin superposition where the component states are spin-$z$ eigenstates. 
In order to generalize it to arbitrary directions and arbitrary number $N$ of spin systems, we use
the angular momentum operator with an arbitrary direction: $ \hat{L}_{\boldsymbol \alpha}=\alpha_x \hat{L}_x + \alpha_y \hat{L}_y + \alpha_z \hat{L}_z$, where $\hat{L}_x=i\sin\phi \ \partial_\theta+i \cot\theta \cos\phi \ \partial_\phi$, $\hat{L}_y=-i\cos\phi \ \partial_\theta+i \cot\theta \sin\phi \ \partial_\phi$ and $\boldsymbol{\alpha}=(\alpha_x,\alpha_y,\alpha_z)$ is a unit vector with a condition $\Vert {\boldsymbol \alpha} \Vert^2={\alpha_x^2+\alpha_y^2+\alpha_z^2}=1$. The multi-spin extension of the definition is
\begin{align}
\mathcal{I}_{\{\boldsymbol \alpha^{(i)}\}} = \frac{1}{2NS\mathcal{P}} \Big(\frac{2S+1}{4\pi}\Big)^N
\int \! d{\bf \Omega} \  W\!\left(\{{\bf n}_i\}\right)  \hat{L}_{\{\boldsymbol \alpha^{(i)}\}}^2 W\!\left(\{{\bf n}_i\}\right)
\label{eq:spin-I-def}
\end{align}
where $d{\bf \Omega} = d\Omega_1 d\Omega_2\cdots d\Omega_N$ and $\{{\bf n}_i\}$ denotes a set of unit vectors $\{ {\bf n}_1,\cdots,{\bf n}_N \}$ and
$\hat{L}_{\{\boldsymbol \alpha^{(i)}\}} = \sum_i \hat{L}_{{\boldsymbol \alpha}^{(i)}}$
is the operator used to capture interference  patterns of the total system
with many spins, where each local operator is oriented in the direction of $\boldsymbol{\alpha}^{(i)}$.
The additional factor $N^{-1}$ was introduced in order to eliminate accumulative microscopic quantum effects according to the size of the system.
We then need to find the optimal directions of $\boldsymbol{\alpha}^{(i)}$ for each local $i$-th spin, which maximizes $\mathcal{I}_{\{\boldsymbol \alpha^{(i)}\}}$.
For the examples illustrated in Fig.~\ref{fig:spin-Wig}, the optimal direction of $\boldsymbol{\alpha}$ is analytically identified to be the $z$-direction. Now, we have a formal definition of the degree of quantum macroscopicity for a system composed of an arbitrary number of spin-$S$ particles as
\begin{align}
\mathcal{I}(\rho) 
= \max_{\{{\boldsymbol \alpha^{(i)}}\}} \mathcal{I}_{\{\boldsymbol \alpha^{(i)}\}}
= \frac{1}{NS\mathcal{P}}\max_{A}\tn{Tr}\left[\rho^2 A^2 - \rho A \rho A \right].
\label{eq:spin-I-def2}
\end{align}
where $A=\sum_{j=1}^N A^{(j)}$ and $A^{(j)}={\boldsymbol \alpha}^{(j)}\!\cdot \hat{{\bf S}}^{(j)}$ is the spin operator.
The upper bound of the measure is found to be $\mathcal{I}(\rho) =NS$
(see the Appendix for the proof). For example, the $N$-party spin-$S$ Greenberger-Horne-Zeilinger (GHZ) state \cite{ghz}, $(\ket{S,S}^{\otimes N} + \ket{S,-S}^{\otimes N})/\sqrt{2}$, has the maximum value ${\cal I}(\rho) = NS$, where the optimal direction of $\boldsymbol{\alpha}^{(i)}$ is the $z$ direction regardless of  $i$.
On the other hand, ${\cal I}(\rho) = S$ for $\left[(\ket{S,S} + \ket{S,-S})/\sqrt{2})\right]^{\otimes N}$
regardless of the number of particles $N$; accumulations of microscopic quantum effects do not increase the value of ${\cal I}(\rho)$.
A spin-$1/2$ GHZ state of $N$ particles, $(\ket{0}^{\otimes N}+\ket{1}^{\otimes N})/\sqrt{2}$, 
has the same value of  ${\cal I}(\rho)=N/2$ with a superposition of a single spin-$N/2$ particle, $(\ket{N/2,N/2}+\ket{N/2,-N/2})/\sqrt{2}$.  Here, we used a common notation $\ket{0} = \ket{1/2,1/2}$ and $\ket{1}=\ket{1/2,-1/2}$ for spin-$1/2$ systems.

If a quantum state is pure, ${\cal I}(\rho)$ reduces to the (normalized) variance of the total spin operator as $\mathcal{I}=\max_{A}\mathcal{V}(A)/(NS)$, where 
$\mathcal{V}(A)=\langle {A^{2}} \rangle - \langle A \rangle ^2$.
Since a macroscopic quantum superposition has well-separated component states of the outcome spectrum, it is in agreement with our natural expectation.
Of course, the variance itself does not allow one to discriminate between a genuine superposition and a statistical mixture.

Another interesting point is  that $\mathcal{I}(\rho)$ quantifies the maximum fragility of a quantum state under dephasing described by $A$.
When a quantum state is subject to a standard Lindblad type of decoherence channel, the decay rate of the logarithm of the purity is given by
\begin{align}
&-\frac{1}{2NS}  \frac{d}{dt}\ln\Tr[\rho^2] 
= \frac{1}{2NS} \left\{-\frac{\mathcal{\dot{P}}}{\mathcal{P}}\right\}\nonumber\\
&=\frac{1}{NS\mathcal{P}}\left\{-\tn{Tr}\left[\rho\mathcal{L}(\rho)\right] \right\} \nonumber\\
&=\frac{\gamma}{NS\mathcal{P}}\tn{Tr}\left[\rho^2 A^2 - \rho A \rho A \right] 
\end{align}
where the Lindblad type of decoherence channel is given by
\begin{align}
	\mathcal{L}(\rho)=\frac{d\rho}{d\tau}=\gamma \Bigl[ A\rho A^\dagger - \frac{1}{2}\left(A^\dagger A \rho + \rho A^\dagger A \right) \Bigr]
\nonumber
\end{align}
where $\gamma$ represents the coupling strength between the system and the environment. 
Using the above expression, we can express ${\cal I}(\rho)$ as 
\begin{align}
	\mathcal{I}(\rho) = -\frac{1}{2NS} \frac{1}{\gamma} \max_A \frac{d}{dt}\ln\Tr[\rho^2] \label{eq:spin-I-sens}.
\end{align}

Measure $\mathcal{I}(\rho)$ can then be understood as fragility of a quantum state. If $\mathcal{I}(\rho)$ is an increasing function of $N$ that diverges to infinity (or at least reaches a large value), $\rho$ becomes extremely fragile for a sufficiently large $N$ regardless of the coupling strength between the system and environment. This is an anomalous situation for a classical system as discussed in Ref.~\cite{Shimizu2002}, and it implies that as far as $\mathcal{I}(\rho)$ is an increasing (and diverging) function of $N$ [for example, ${\cal I}(\rho)=N^\epsilon S$  with $\epsilon>0$], the state becomes a macroscopic superposition as $N$ increases. 
This property is consistent with that of the original definition of ${\cal I}(\rho)$ for harmonic oscillator systems~\cite{LeeJeong2011}.

We here note that $\mathcal{I}(\rho)$ is not convex under classical mixing. This is due to the purity in the denominator in Eq.~\eqref{eq:spin-I-def2} added to its original definition in Ref.~\cite{LeeJeong2011} for harmonic oscillator systems. 
For example, let us consider a mixture of one Bell state and the fully decohered state $\rho = a \rho_0 + (1-a) \rho_1$ where $\rho_0=(\ket{00}\bra{00}+\ket{11}\bra{11})/2$ and $\rho_1=(\ket{00}+\ket{11})(\tn{h.c.})/2$. For $a=0,1/2$, and $1$, the values of $\mathcal{I}(\rho)$ are $1$, $9/10$, and $1/2$, respectively.

Hereafter, we focus on quantum macroscopicity of $N$-partite qubit (i.e. $S=1/2$) states.
For simplicity, we normalize the maximum value of $\cal I(\rho)$ for any state $\rho$ of $N$ spin-1/2 particles to be  $N$ rather than $NS=N/2$, which can be done simply by multiplying $2$ to the original definition of $\mathcal{I}(\rho)$ in Eq.~(\ref{eq:spin-I-def2}).

\subsection{Comparison with QFI based measure}
Although our measure and the QFI based one \cite{Frowis2012NJP} are devised from different starting points, they have similar mathematical structures.
The QFI is defined as $F(\rho,A)=2 \sum_{i,j=1}^{2^N} {(\pi_i-\pi_j)^2}/{(\pi_i+\pi_j)}|\bra{i}A \ket{j}|^2$,
where $\pi_i$($\ket{i}$) is $i$th eigenvalue(eigenvector) of the density matrix $\rho$ and 
$A=\sum_{j=1}^N {\boldsymbol \alpha}^{(j)}\cdot {{\boldsymbol \sigma}}^{(j)}$ with Pauli operators ${\boldsymbol \sigma}^{(j)}$ for the $j$th site with  $\Vert{\boldsymbol \alpha}^{(j)}\cdot {{\boldsymbol \sigma}}^{(j)}\Vert^2=\Vert{\boldsymbol \alpha}^{(j)}\Vert^2=1$.
The effective size of a macroscopic quantum state is then defined as \cite{Frowis2012NJP}
\begin{align}
\mathcal{F}(\rho)\equiv \max_{A}\frac{F(\rho,A)}{4N}=\frac{1}{2N}\max_{A}\sum_{i,j=1}^{2^N} \frac{(\pi_i-\pi_j)^2}{(\pi_i+\pi_j)}|\bra{i}A
\ket{j}|^2,
\label{eq:fi}
\end{align}
and it has the maximum value ${\cal F}(\rho)=N$ for an $N$-partite GHZ state.
Using Eq.~(\ref{eq:spin-I-def2}), we can rewrite $\mathcal{I}(\rho)$ (with an extra normalization factor $2$ mentioned above) as 
\begin{align}
\mathcal{I}(\rho)= \frac{1}{2N}\max_{A} \sum_{i,j=1}^{2^N}\frac{(\pi_i-\pi_j)^2}{\sum_k{\pi_k^2}}\abs{\bra{i}A\ket{j}}^2.
\end{align}
It is clear that the only difference between $\mathcal{I}(\rho)$ and $\mathcal{F}(\rho)$ is the denominator of the weights of $\abs{\langle i | A | j \rangle}^2$ which are $\sum_k{\pi_k^2}$ and $\pi_i+\pi_j$, respectively.
Both $\mathcal{I}(\rho)$ and $\mathcal{F}(\rho)$ become identical to the maximum variance {\it per} particle, $\max_A\mathcal{V}(A)/N$, for any pure state.

For pure states, this measure has been compared with multipartite entanglement~\cite{yadin15,tichy16}. The results show that even if a quantum state has a certain type of multipartite entanglement, it does not necessarily mean that the state is a macroscopic superposition. For instance, any two spins comprising a cluster state have localizable entanglement but the cluster state only gives $\mathcal{I}=\mathcal{F}=O(1)$ which does not indicate quantum macroscopicity~\cite{yadin15}. A Haar random state in many-particle Hilbert space is another example which has a large degree of geometric entanglement but is not a macroscopic superposition~\cite{tichy16}.

We compare time complexities of numerical methods to obtain $\mathcal{I}(\rho)$ and $\mathcal{F}(\rho)$ in Appendix~\ref{sec:time_compare}.
In accordance with our expectation, the results show that $\mathcal{I}$ can be calculated in $O(D^2)$ operations while $O(D^3)$ operations are required for $\mathcal{F}$ where $D=2^N$ is the dimension of the density matrix.

\subsection{Examples for mixed states}
As an example, consider a generalized mixed GHZ state
\begin{align}
\rho_{\tn G}= \mathcal{N}^{-1}\left( \pjct{0}{0}^{\otimes N} \!+ \pjct{\epsilon}{\epsilon}^{\otimes N}\!+\gamma \pjct{0}{\epsilon}^{\otimes N}\!+\gamma\pjct{\epsilon}{0}^{\otimes N}\right) 
\label{eq:Dur}
\end{align}
where $\ket{\epsilon} = \cos{\epsilon}\ket{0} + \sin{\epsilon} \ket{1}$ and $\mathcal{N}=2(1+\gamma\cos^N\epsilon)$.	
The two components of the superposition, $\ket{0}^{\otimes N}$ and $\ket{\epsilon}^{\otimes N}$, have an overlap of $\langle 0\vert\epsilon\rangle^N=\cos^N{\epsilon}$.
For this state, we can analytically calculate $\mathcal{I}$ and $\mathcal{F}$ using Eqs.~\eqref{eq:spin-I-def2} and ~\eqref{eq:fi}, respectively. 
The two measures give the same value in the limit $\epsilon\ll1$ and $N \gg 1$, which is $\mathcal{I}\simeq\mathcal{F}\approx \gamma N \epsilon^2/(1+\gamma) + O(1)$.
The same value of the measures $\mathcal{I}=\mathcal{F}$ is also obtained for $\gamma=1$, when $\rho_{\tn G}$ is a pure state.
The ratio between two measures $\mathcal{I}(\rho_{\tn G})/\mathcal{F}(\rho_{\tn G})$ has the maximum value $2$ in the limit of a large system ($N\gg1$) with a high mixture ($\gamma \ll 1$) and large distinguishably between the component states ($\epsilon=\pi/2$).

\begin{figure}[t!]
\centering
	\resizebox{0.4\textwidth}{!}{\includegraphics{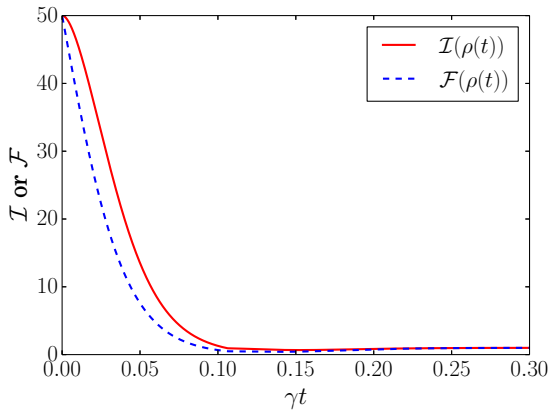}}
	\caption{
		(Color online) For the initial GHZ state with $N=50$, we calculated the dynamics of $\mathcal{I}(\rho)$ and $\mathcal{F}(\rho)$ under the dissipation given by master equation Eq.~\eqref{eq:master-eq}.
}
\label{fig:ghz}
\end{figure}

\begin{figure*}[t!]
	\centering
	\resizebox{0.7\textwidth}{!}{\includegraphics{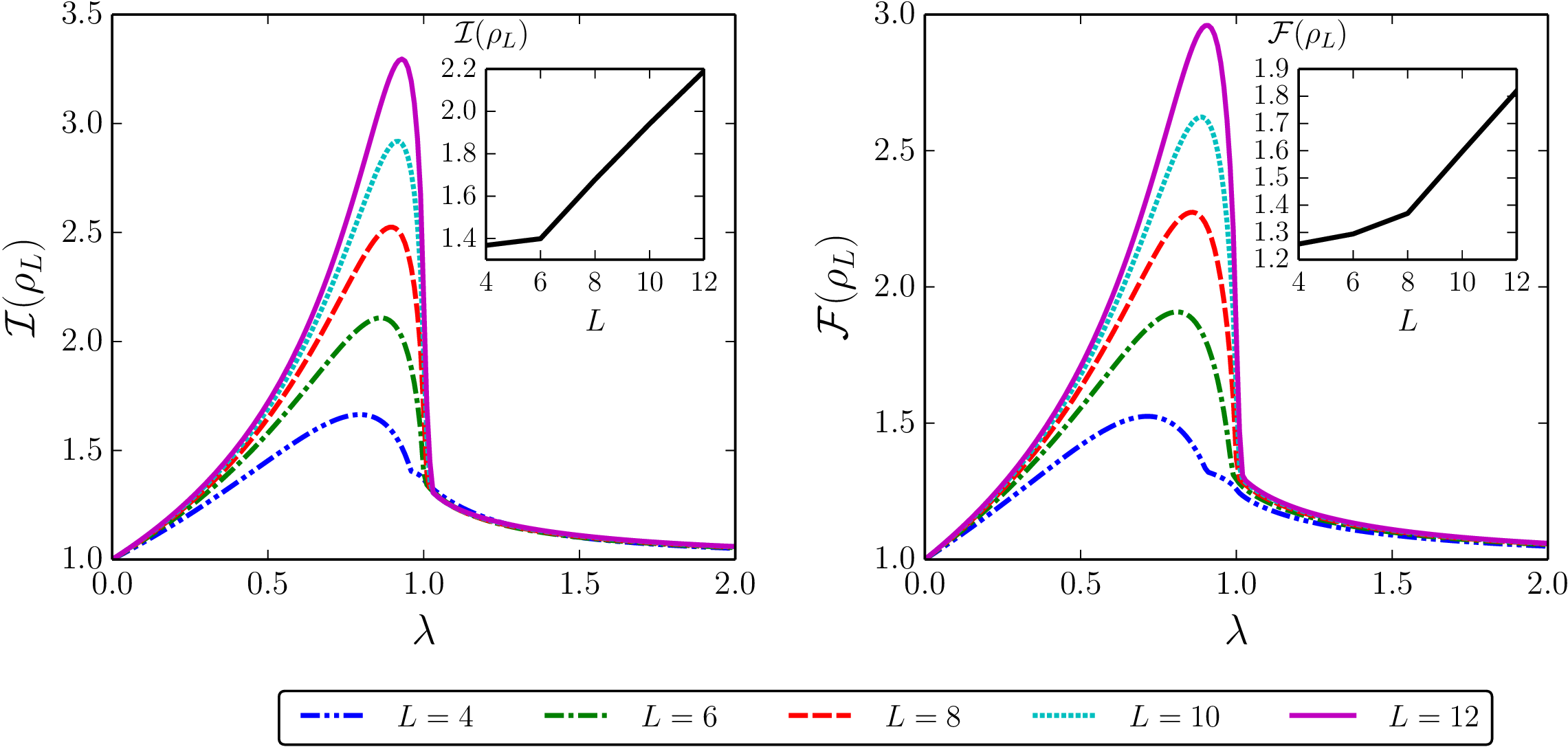}}
	\caption{
		(Color online) Quantum macroscopicity in terms of $\mathcal{I}(\rho_L)$ (left) and $\mathcal{F}(\rho_L)$ (right) for a partial block of $L$ contiguous particles in the ground state versus the interaction strength $\lambda$ of the transverse Ising model. The QPT occurs at the critical point $\lambda=1$.
(Insets) Values of each measure as a function of $L$ at the critical point of $\lambda=1$.
}
\label{fig:i-lambda}
\end{figure*}

We investigate another type of mixed state that shows  sub-optimal precision for quantum metrology \cite{KavanPRX}
\begin{align}
\rho_{\tn{M}}
=\frac{1}{2}\left(
\begin{array}{cc}
\rho_0^{\otimes N-1} & p(\rho_0 \sigma_x)^{\otimes N-1} \\
 p(\sigma_x \rho_0)^{\otimes N-1} &  (\sigma_x \rho_0 \sigma_x)^{\otimes N-1}
\end{array}
\right),
\label{eq:rho-metrology}
\end{align}
where $\rho_0=\{(1+p)\pjct{0}{0}+(1-p)\pjct{1}{1}\}/2$.
It is straightforward to obtain 
$\mathcal{I}{(\rho_{\tn{M}})}=8p^4 N/(1+p^2)^3  +O(1)$.
However, it is necessary to calculate the eigenvalues and eigenvectors of $\rho_M$ in order to
obtain its QFI, which was done in Ref.~\cite{KavanPRX} and
we thus get $\mathcal{F}{(\rho_{\tn{M}})}=p^4 N + O(1)$.
This is a typical example that shows the computational advantage of $\mathcal{I}(\rho)$.
The ratio between the two measures becomes $\mathcal{I}(\rho_{\tn{M}})/\mathcal{F}(\rho_{\tn{M}}) \simeq 8$ for $p \ll 1$ and $N\gg1$, and this  is even larger compared to the previous example.

\subsection{GHZ state under dissipation}
Even though the formal definition of the measure $\mathcal{I}(\rho)$ is related to the sensitivity of a quantum state under dephasing [Eq.~\eqref{eq:spin-I-sens}], $\mathcal{I}(\rho)$ also well captures the dynamics of a quantum state under dissipation. For example, we consider the master equation which is given by
\begin{align}
\frac{\partial \rho}{\partial t} = -i \frac{\Omega}{2}[J_+ + J_-, \rho] + \frac{\gamma}{2} \bigl[ 2 J_{-} \rho J_{+} - J_{+} J_{-}\rho- \rho J_{+} J_{-} \bigr] \label{eq:master-eq}
\end{align}
where $J_{+}=\sum_{i=1}^N \sigma^{(i)}_+ $ and $J_=\sum_{i=1}^N \sigma^{(i)}_{-}$ are the collective spin raising and lowering operators, respectively, where $\sigma^{(i)}_\pm = (\sigma^{(i)}_x \pm i \sigma^{(i)}_x)/2$ are the single particle spin operators. In addition, $\Omega$ is the Rabi frequency and $\gamma$ is the dissipation rate.
This master equation can be derived from the Dicke model Hamiltonian with appropriate approximations of a bosonic bath mode~\cite{agarwal74,drummond78}.
We here consider the limit $\Omega \ll \gamma$, and therefore ignore the first term in the master equation. In this limit and for the initial GHZ state with $N=50$, the dynamics of $\mathcal{I}[\rho(t)]$ and $\mathcal{F}[\rho(t)]$ is numerically calculated and plotted in Fig.~\ref{fig:ghz}. At first, for $0 \leq \gamma t \lesssim 0.14$, the measures first decrease to the values smaller than $1$ as time increases. After that, the values slightly increase to $1$ as $\gamma t \rightarrow \infty$. This is because the quantum state $\rho(t)$ becomes a highly mixed state with low quantum coherence by the dissipation at first, but evolves into the pure steady state $\rho_\infty = \ket{1}\bra{1}^{\otimes N}$ which has the value $\mathcal{I}(\rho_\infty)=\mathcal{F}(\rho_\infty)=1$. This is the minimum value of the measures among pure states. This example well demonstrates the loss of macroscopic quantum coherence of the initial GHZ state by dissipation and the nature of a steady state as a pure state.

\section{Application to quantum phase transition.}
The QPT is a well known quantum effect where the ground state of a many-body system experiences an abrupt change of phase,
which is characterized by the change of an observable called an order parameter, with the change of an external parameter. 
The two distinct phases with different values of an order parameter have different macroscopic properties, and the quantum coherence can be preserved at $T=0$. We thus expect that macroscopic quantum coherence between the two distinct states may appear at the critical point where QPT occurs.

We investigate this conjecture using the measures $\mathcal{I}$ and $\mathcal{F}$. There have been many studies which revealed the relation between QPT and entanglement~\cite{Osborne2002, Vidal2003,Osterloh2002}. However, it does not directly show the possibility of macroscopic quantum superposition during QPT as noted above.
The transverse Ising model is the simplest quantum many-body model exhibiting QPT~\cite{sachdevQPT}.
Its Hamiltonian is
\begin{align}
H_{\tn{Ising}}(\lambda)=-\sum_{j=1}^{N} \left( \lambda \sigma_x^{(j)}\sigma_x^{(j+1)} + \sigma_z^{(j)}\right)
\label{eq:Ising-Hamiltonian}
\end{align}
where $\sigma_i^{(j)}$ is the $i$-component Pauli operator on the  $j$-th site, $\lambda$ is the interaction strength, and  a periodic boundary condition ${\boldsymbol \sigma}^{(N+1)}\equiv{\boldsymbol \sigma}^{(1)}$ is given.
The QPT then occurs at the critical point $\lambda=\lambda_c=1$ where the global phase-flip symmetry of the ground state is broken \cite{Osborne2002}.
For $\lambda<\lambda_c$, the ground state is in the disordered paramagnetic phase with the order parameter $\langle\sigma_x\rangle=0$ while it becomes the ordered ferromagnetic phase for $\lambda>\lambda_c$ with $\langle\sigma_x\rangle\neq0$.

First, we simply investigate the behavior of the measures, $\cal I$ and $\cal F$, as function of $N$ at the critical point of QPT for the ground state of a finite spin chain.
For pure states, both $\cal I$ and $\cal F$ give the same value which is the maximum variance of $A$ divided by $N$. The variance of $A$ in terms of spin-spin correlations can be expressed as
\begin{align*}
\braket{A^2}-\braket{A}^2 =& \,N-\sum_{i=1}^N \braket{A^{(i)}}^2 + \\
	&\sum_{i\neq j} \bigl[ \braket{A^{(i)} A^{(j)}} - \braket{A^{(i)}}\braket{A^{(j)}} \bigr],
\end{align*}
noting that $(A^{(i)})^2=\openone$ with $A^{(i)}={\boldsymbol \alpha}^{(i)}\!\cdot {\boldsymbol \sigma}^{(i)}$. Therefore, if the spin-spin correlation function,
$\braket{A^{(i)} A^{(j)}} - \braket{A^{(i)}}\braket{A^{(j)}}$, decays faster than $O(1/|i-j|)$, the variance is $O(N)$ and the measures will give values of $O(1)$. 
However, if the correlation function is $O(1/|i-j|^\gamma)$ for $\gamma<1$, the variance 
will be $O(N^{2-\gamma})$. 
It is known that the spin-spin correlation function along the $x$ direction, $\braket{\sigma^{(i)}_x\sigma^{(j)}_x}-\braket{\sigma^{(i)}_x}\braket{\sigma^{(j)}_x}$, decays as $O(1/|i-j|^{1/4})$~\cite{pfeuty70} at the critical point in the thermodynamic limit ($N \rightarrow \infty$), which implies $\mathcal{I}=\mathcal{F}\geq O(N^{3/4})$. This indicates the existence of quantum macroscopicity at the critical point of the QPT. 
As this correlation function is derived in the thermodynamic limit, it is unclear whether it is correct also for finite values of $N$. In fact, to make a detailed analysis, one needs to appropriately deal with the finite size effect using the universal behavior of physical variables that arises near the critical point. References~\cite{hauke16,abad16} performed this type of analysis and their results are consistent with ours, i.e., quantum Fisher information (the maximum variance for pure states) of the order of $O(N^{3/4})$.

As a crucial property of a critical system is its scale invariance, we need to investigate whether a finite subsystem of the entire system shows a similar behavior. 
We take the thermodynamic limit of the model (i.e. $N\rightarrow\infty$) and consider a partial block of length $L$ contiguous particles of the ground state.
A subsystem $\rho_L$ of  block size $L$ is generally not a pure state because $L$ spins are correlated with the rest part of the spin chain.
In order to obtain $\mathcal{I}(\rho_L)$ and $\mathcal{F}(\rho_L)$, 
we first find explicit forms of $\rho_L$ and $A$ in terms of the Pauli matrices for a number of given block sizes up to $L=12$ using the analytic solution of $\rho_L$ obtained in Ref.~\cite{Vidal2003} (see Appendix \ref{sec:transverseising}).
We put those forms into Eqs.~\eqref{eq:spin-I-def2} and~\eqref{eq:fi},
and perform the numerical optimization procedure over all possible directions of $\boldsymbol{\alpha}^{(i)}$ for each site $i$.

The result is shown in Fig.~\ref{fig:i-lambda} for block sizes up to $L=12$.
When $\lambda\rightarrow0$, the ground state is $\ket{\uparrow}^{\otimes N}$ and any subsystem is a product state with $\mathcal{F}(\rho_L)=\mathcal{I}(\rho_L)=1$. When $\lambda\rightarrow\infty$, 
a subsystem of the symmetry preserving ground state is $\rho_L=(\ket{\!\rightarrow}\bra{\rightarrow\!})^{\otimes L} + (\ket{\!\leftarrow}\bra{\leftarrow\!})^{\otimes L}$  
and a simple calculation yields the same value $\mathcal{I}(\rho_L)=\mathcal{F}(\rho_L)=1$. 
As $L$ increases, we observe that $\mathcal{I}(\rho_L)$ peaks right before the critical point $\lambda_c=1$ and rapidly decreases as the interaction becomes stronger.
This result shows that the subsystems of the spin chain near the quantum phase transition have quantum macroscopicity. This is an interesting result in comparison to typical examples of multipartite macroscopic superpositions such as GHZ states~\cite{ghz}, NOON states~\cite{noon}, and entangled coherent states~\cite{ecs} of which the subsystems have no quantum macroscopicity.

The above arguments imply that an arbitrarily large macroscopic quantum superposition may arise during QPT of a large transverse Ising system.
Our result suggests that the QPT may be regarded as a genuine macroscopic quantum phenomenon in which the quantum coherence between macroscopically distinct states is involved.

%\textit{Remarks.}--
\section{Remarks}
In summary, we have suggested a general %and computable
 measure of quantum macroscopicity for arbitrary spin systems,
and found that a macroscopic superposition of an extremely large size arises during the QPT of the transverse Ising model near the critical point.
Although the transverse Ising model is a paradigmatic example, it is worth investigating other models.
How to choose criteria of a good macroscopicity measure 
\cite{Yadin2016} is another topic that deserves rigorous investigations.
Our study provides new insight into a classic many-body phenomenon in relation to the notion of quantum macroscopicity, and this may be an important step  towards deeper understanding of quantum macroscopicity of many-body %physical 
systems.

\section{Acknowledgement }
This work was supported by the National Research Foundation of Korea (NRF) grant funded by the Korea government (MSIP) (No. 2010-0018295) and by the KIST Institutional Program (Project No. 2E26680-16-P025).

\appendix

\section{Purity expressed in terms of the characteristic function}
An explicit form of the Wigner distribution for a spin-$S$ particle is \cite{GroupTextbook}
\begin{equation}
W({\bf n}) =\sqrt{\frac{4\pi}{2S+1}}\sum_{L=0}^{2S}\sum_{M=-L}^{L} \chi_{L,M}^{(S)} \  Y_{L,M}({\bf n}),
\end{equation}
where $Y_{L,M}({\bf n})$ denotes spherical harmonics with a three-dimensional unit vector ${\bf n}$=(sin$\theta$cos$\phi$, sin$\theta$sin$\phi$, cos$\theta$) and $\chi_{L,M}^{(S)}=\Tr[\hat{T}_{L,M}^{(S)\dagger}\rho ]$ is the characteristic function with the irreducible tensor operator $\hat{T}_{L,M}^{(S)}$ defined in the main text and the density operator $\rho$.
The Wigner distribution  can also be expressed as
$\Tr\left[\rho \hat{w}({\bf n})\right]$ with the transformation kernel \cite{GroupTextbook}
\begin{equation}
\hat{w}({\bf n}) = \sqrt{\frac{4\pi}{2S+1}}\sum_{L=0}^{2S}\sum_{M=-L}^{L} \hat{T}_{L,M}^{(S)\dagger} Y_{LM}({\bf n}).
\end{equation}
In general, the $W$ symbol of an arbitrary operator $\hat{f}$ is defined as $W_f({\bf n})=\Tr\left[ \hat{f}\hat{w}({\bf n})\right]$ and the expectation value of $\hat{f}$ can be obtained by
\begin{equation}
\Tr\left[\rho \hat{f}\right] = \frac{2S+1}{4\pi} \int  d\Omega \  W({\bf n}) W_f({\bf n}).
\label{eq:W-innerproduct}
\end{equation}
Using the orthogonality of the spherical harmonics, 
\begin{equation}
\int d\Omega Y_{L,M} ({\bf n})Y^*_{L',M'}({\bf n}) = \delta_{L,L'}\delta_{M,M'},
\end{equation}
we can inversely obtain the characteristic function from the Wigner distribution as
\begin{align}
\sqrt{\frac{2S+1}{4\pi}} &\int d\Omega Y^*_{L,M}({\bf n}) W({\bf n}) \nonumber \\
&= \sum_{L',M'}  \int d\Omega Y_{L',M'}({\bf n}) Y^*_{L,M}({\bf n}) \chi_{L',M'}^{(S)} \nonumber \\
&= \sum_{L',M'}   \delta_{L,L'}\delta_{M,M'} \chi_{L',M'}^{(S)} \nonumber \\
&= \chi_{L,M}^{(S)}.
\label{eq:char-in-Wig}
\end{align}
We then find a relation between the purity $\mathcal{P}(\rho)$ of  state $\rho$ and the characteristic function $\chi_{L,M}^{(S)}$ as
\begin{align}
&\sum_{L,M} \abs{\chi_{L,M}^{(S)}}^2 \nonumber \\
&=\frac{2S+1}{4\pi}\! \int \! d\Omega d\Omega' \left(\sum_{L,M} Y_{L,M}({\bf n}) Y^*_{L,M}({\bf n'})\right)\!\!  W({\bf n}) W({\bf n'}) \nonumber \\
&=\frac{2S+1}{4\pi} \int  d\Omega d\Omega' \delta({\bf n}-{\bf n'})  W({\bf n}) W({\bf n'}) \nonumber \\
&=\frac{2S+1}{4\pi} \int  d\Omega W^2({\bf n}) 
%\nonumber \\ &
=\Tr[\rho^2] = \mathcal{P}(\rho).
\end{align}

\section{Wigner representation of the measure of quantum macroscopicity}

We can prove Eq.~\eqref{eq:i-guess-wig} in the main text as follow. Using Eq.~(\ref{eq:char-in-Wig}) in the previous section and the theorem of integration by parts, we find
\begin{align}
&\mathcal{I}_z=\frac{1}{2S\mathcal{P}}\sum_{L=0}^{2S} \sum_{M=-L}^{L} M^2 \abs{\chi_{ L, M}^{(S)} }^2 \nonumber \\
&=\frac{1}{2S\mathcal{P}}\sum_{L=0}^{2S} \sum_{M=-L}^{L} M^2  {\chi_{ L, M}^{(S)*}}{\chi_{ L, M}^{(S)}}  \nonumber \\
&=\frac{1}{2S\mathcal{P}}\sqrt{\frac{2S+1}{4\pi}}\! \sum_{L,M} \int \! d\Omega M^2 Y_{L,M}({\bf n}) W({\bf n}) \chi_{L,M}^{(S)} \nonumber \\
&=-\frac{1}{2S\mathcal{P}}\sqrt{\frac{2S+1}{4\pi}}\! \sum_{L,M} \int \! d\Omega\frac{\partial^2Y_{L,M}({\bf n})}{\partial\phi^2} W({\bf n}) \chi_{L,M}^{(S)}\nonumber \\
&=-\frac{2S+1}{8\pi S\mathcal{P}}\! \sum_{L,M} \int \! d\Omega d\Omega' Y_{L,M}({\bf n}) \frac{\partial^2W({\bf n}) }{\partial\phi^2}  Y^*_{L,M}({\bf n'})   W({\bf n'}) \nonumber \\
&=-\frac{2S+1}{8\pi S\mathcal{P}}\! \int \! d\Omega d\Omega'  \delta({\bf n}-{\bf n'})  \frac{\partial^2W({\bf n}) }{\partial\phi^2} W({\bf n'}) \nonumber \\
&=-\frac{2S+1}{8\pi S\mathcal{P}}\! \int \! d\Omega \frac{\partial^2W({\bf n}) }{\partial\phi^2} W({\bf n})\nonumber \\
&=\frac{2S+1}{8\pi S\mathcal{P}}\int d\Omega \ W({\bf n})\hat{L}_z^2 W({\bf n}),
\end{align}
so that the two lines of  Eq.~\eqref{eq:i-guess-wig} in the main text are identical.

\section{ Density operator representation of the measure of quantum macroscopicity}

We now show that the two definitions of $\mathcal{I}_{\boldsymbol{\alpha}}$, Eq.~\eqref{eq:spin-I-def} of the main text and the objective function that undergoes the maximization in Eq.~\eqref{eq:spin-I-def2} of the main text, for a single particle are equivalent.
The density matrix of $\mathcal{I}_{\boldsymbol{\alpha}}$ can be expressed as
\begin{align}
\mathcal{I}_{\boldsymbol{\alpha}} &= \frac{1}{S\mathcal{P}} \left(\Tr\left[\rho^2 A^2\right]-\Tr\left[\rho A \rho A \right] \right)\nonumber\\
&= \frac{1}{S\mathcal{P}}\sum_{i,j=x,y,z} \alpha_i \alpha_j \left(\Tr[\rho^2 S_i S_j] - \Tr[\rho S_i \rho S_j ]\right) \nonumber \\
&= \frac{1}{S\mathcal{P}}\sum_{i,j=x,y,z} \alpha_i \alpha_j  \Tr [\rho \hat{f}_{ij}],
\label{eq:I-def-single}
\end{align}
where $\hat{f}_{ij}=([S_i,S_j\rho]+[\rho S_i,S_j])/2$
and  $A={\boldsymbol \alpha}\!\cdot \hat{{\bf S}}$ with unit vector ${\boldsymbol \alpha}$.
Since the trace of any two operators can be calculated as Eq.~(\ref{eq:W-innerproduct}), we need to obtain the $W$ symbol of $\hat{f}_{ij}$ as
\begin{align}
W_{f_{ij}}=\Tr[\hat{f}_{ij}\hat{w}({\bf n})] & =\frac{1}{2} \Tr[S_i S_j \rho \hat{w} + \rho S_i S_j \hat{w} - 2 S_j \rho S_i \hat{w}] \nonumber \\
&=\frac{1}{2} \Tr[\rho \hat{w}  S_i S_j+ \rho S_i S_j \hat{w} - 2  \rho S_i \hat{w}S_j] \nonumber \\
&=\frac{1}{2} \Tr[\rho  ([\hat{w},S_i]S_j+S_i [S_j,\hat{w}])] \nonumber \\
&=\frac{1}{2}\hat{L}_i\Tr[\rho   \hat{w}  S_j ] -\frac{1}{2}  \hat{L}_j\Tr[\rho  S_i \hat{w}], \nonumber \\
\end{align}
where the last equality holds for $[\hat{w},S_i]=\hat{L}_i\hat{w}$, which is proven in Ref.~\cite{Kalmykov2008}.
When $i=j$, it is further simplified as 
\begin{equation}
W_{f_{ii}} = \frac{1}{2} \hat{L}_i \Tr[\rho [\hat{w},S_i]] = \frac{1}{2} \hat{L}_i^2 \Tr[\rho \hat{w}] = \frac{1}{2} \hat{L}_i^2 W({\bf n}).
\label{eq:i=j}
\end{equation}
If $i\neq j$,  $\hat{f}_{ij}$ always has its symmetric pair $\hat{f}_{ji}$ in Eq.~(\ref{eq:I-def-single}) so that we can combine them as
\begin{align}
W_{f_{ij}}+W_{f_{ji}}&=\Tr[(\hat{f_{ij}}+\hat{f}_{ji})\hat{w}] \nonumber \\
&=\frac{1}{2} \hat{L}_i \Tr \left[ \rho [\hat{w},S_j] \right]+ \frac{1}{2} \hat{L}_j \Tr \left[ \rho [\hat{w},S_i] \right] \nonumber \\
&= \frac{1}{2} \left(\hat{L}_i \hat{L}_j+\hat{L}_j \hat{L}_i\right) \Tr[ \rho \hat{w}] \nonumber \\
&= \frac{1}{2} \left(\hat{L}_i \hat{L}_j+\hat{L}_j \hat{L}_i\right) W({\bf n}).
\label{eq:i=!j}
\end{align}
Using Eqs.~(\ref{eq:W-innerproduct}), (\ref{eq:i=j}) and (\ref{eq:i=!j}), Eq.~(\ref{eq:I-def-single}) becomes
\begin{align}
\mathcal{I}_{\boldsymbol{\alpha}} &= \frac{2S+1}{8\pi S\mathcal{P}} \sum_{i,j=x,y,z} \alpha_i \alpha_j \int d\Omega \ W({\bf n}) \hat{L}_i \hat{L}_j W({\bf n}) \nonumber \\
 &= \frac{2S+1}{8\pi S\mathcal{P}} \int  d\Omega \ W({\bf n}) \hat{L}_{\boldsymbol{\alpha}}^2 W({\bf n}).
\end{align}
Its extension to an arbitrary number of spin is straightforward because the measure contains at most the quadratic order of spin operators and the operators for different sites commute.

\section{ Upper bound of the measure of quantum macroscopicity}

We here show that the upper bound of ${\cal I}(\rho)$ for an arbitrary multipartite system $\rho$ composed of $N$ spin-$S$ particles
is $NS$.
%It is straightforward to derive Eq.~(4) from Eq.~(3) as
We use the definition of $\mathcal{I}$ in Eq.~\eqref{eq:spin-I-def2} in the main text,
\begin{align}
&\mathcal{I}(\rho)
=
\frac{1}{NS\mathcal{P}}\max_{A}\tn{Tr}\left[\rho^2 A^2 - \rho A \rho A \right].
\label{eq:I-rho}
\end{align}
 By noting that $\Tr[(\rho A)^2] \geq 0$, we find 
\begin{align}
&\Tr[\rho^2 A^2] - \Tr[\rho A \rho A] \nonumber \\
\leq& \Tr[ \rho^2 A^2]  = \sum_{i=i}^{2^N} \pi_i^2 \langle i \vert A^2 \vert i \rangle =  \sum_{i=i}^{2^N}\pi_i^2\sum_{j,k=1}^N  \langle i \vert A^{(j)} A^{(k)} \vert i \rangle \nonumber \\
%\leq& \Tr[\rho A^2] = \langle A^2 \rangle =\sum_{i,j=1}^N \langle A^{(i)} A^{(j)} \rangle \nonumber \\
\leq& \sum_{i=i}^{2^N}\pi_i^2\sum_{j,k=1}^N  S^2 =  \sum_{i=i}^{2^N}\pi_i^2 N^2  S^2 = \mathcal{P} N^2 S^2,
%\leq&\sum_{i,j=1}^N S^2 =N^2S^2
\label{eq:upperbound}
\end{align}
where $\pi_i$ ($\ket{i}$) is the $i$-th eigenvalue (eigenvector) of the density operator $\rho$.
Using Eqs.~(\ref{eq:I-rho}) and (\ref{eq:upperbound}), we obtain 
\begin{equation}
\mathcal{I}(\rho)\leq \frac{1}{NS\mathcal{P}} \mathcal{P} N^2S^2 = NS.
\end{equation}
As an example, we find that the $N$-partite spin-$S$ GHZ state,
\begin{equation}
\ket{\tn{GHZ}}=\frac{1}{\sqrt{2}}\left(\ket{S,S}^{\otimes N} + \ket{S,-S}^{\otimes N}\right),
\end{equation}
has the maximal value ${\cal I}(\rho)=NS$ when the operator $A^{(i)}$ is aligned along the $z$ direction for every value of $i$.

\section{Time complexities for calculating $\mathcal{I}$ and $\mathcal{F}$} \label{sec:time_compare}

Here, we provide a method to calculate measures $\mathcal{I}$ and $\mathcal{F}$ and compare their time complexities. We here just consider a many-particle spin-$1/2$ case for the convenience but the same argument is directly applicable for any spin-$S$ case.
First, we note that the value of $\mathcal{I}$ for a given $\rho$ can be calculated using a $3N$ by $3N$ matrix $V$ with its components
\begin{align*}
	V_{ia,jb} &= \frac{1}{N\mathcal{P}} \Tr[\rho^2 \sigma^{(i)}_a \sigma^{(j)}_b - \rho \sigma^{(i)}_a \rho \sigma^{(j)}_b]
\end{align*}
where $1\leq i,j \leq N$ are the indices for the spin and $a,b \in \{x,y,z\}$ are the indices for the direction of the spin operator. Using the matrix $V$, the measure $\mathcal{I}$ can be expressed as
\begin{align*}
	\mathcal{I}(\rho) = \max_{\{\alpha^{(i)}_a\}} \sum_{i,j,a,b} \alpha^{(i)}_a \alpha^{(j)}_b V_{ia,jb}.
\end{align*}
In other words, the value of $\mathcal{I}$ can be obtained by solving the following optimization problem,
\begin{equation}
\label{eq:optimize-I}
\begin{aligned}
	& \underset{\pmb{\alpha}}{\text{maximize}} & & \langle \pmb{\alpha} , V\pmb{\alpha} \rangle \\
	& \text{subject to} 	& & |\pmb{\alpha}^{(i)}| = 1, \; i=1,\ldots,N. 
\end{aligned}
\end{equation}
where $\pmb{\alpha} = \{\alpha^{(1)}_x,\alpha^{(1)}_y,\alpha^{(1)}_z,\cdots,\alpha^{(N)}_x,\alpha^{(N)}_y,\alpha^{(N)}_z\}$ is a $3N$ component vector and $\langle \cdot, \cdot \rangle$ is the inner product between two vectors. 

Similarly, the measure $\mathcal{F}$ can be calculated using a matrix $W$, which is defined as
\begin{align*}
	W_{ia,jb} = \frac{1}{2N} \sum_{k,l=1}^{2^N} \frac{(\pi_k - \pi_l)^2}{\pi_k+\pi_l} \braket{k|\sigma^{(i)}_a|l}\braket{l|\sigma^{(j)}_b|k}
\end{align*}
where $\pi_i$ and $\ket{i}$ are the $i$-th eigenvalue and eigenstate of a density matrix $\rho$, respectively. By replacing the matrix $V$ in the optimization problem~\eqref{eq:optimize-I} with $W$, we can calculate the optimized value of $\mathcal{F}$ for the quantum state $\rho$.

The above argument shows that the optimization problems for $\mathcal{I}$ and $\mathcal{F}$ are in the same class.
Therefore, the difference between the calculation times for $\mathcal{F}$ and $\mathcal{I}$ mainly comes from the construction of the matrices $V$ and $W$ as they requires .
Then, let us compare the number of operations needed to construct the matrices $V$ and $W$. First, using $\Tr[\rho^2\sigma^{(i)}_a\sigma^{(j)}_b] = \Tr[\rho\sigma^{(i)}_a (\rho\sigma^{(j)}_b)^\dagger]$ and the sparseness of $\sigma^{(i)}_a$, we can calculate each component of matrix $V$ in $O(D^2)$ operations where $D=2^N$ is the dimension of the density matrix $\rho$. 
On the other hand, for the calculation of $W$, we first need to diagonalize $\rho$ which usually requires $O(D^3)$ operations. Moreover, $O(D^2)$ operations are additionally required to calculate each component of $W$ from the summation. In summary, the number of operations to construct $W$ is $O(D) \gg 1$ times larger than that of $V$.

Then how much time would be consumed for the optimizations? 
If the time for optimization dominates the time to construct the matrices, the total time complexities for $\mathcal{I}$ and $\mathcal{F}$ may not differ much.
In fact, our optimization problem is an quadratically constrained quadratic program which is generally NP-hard as 0-1 integer programming which is NP-hard can be converted into this form. As the number of optimizing variables is $3N$ and the number of constraints is $N$ for our problem, more than exponential time of $N$ is required to obtain the complete solution of the optimization problem.
Nevertheless, typical numerical optimization problems including ours only require \textit{practically approvable} solutions not the complete solutions of the problem.
Many algorithms for function optimization are known to give such feasible solutions in a polynomial time of $N$ for this problem. 
Thus, we can expect that the optimization time which is polynomial in $N$ does not contribute much to the total time for calculations which is polynomial in $D=2^N$.

\begin{figure}[t]
	\centering
	\resizebox{0.48\textwidth}{!}
	{
		\includegraphics{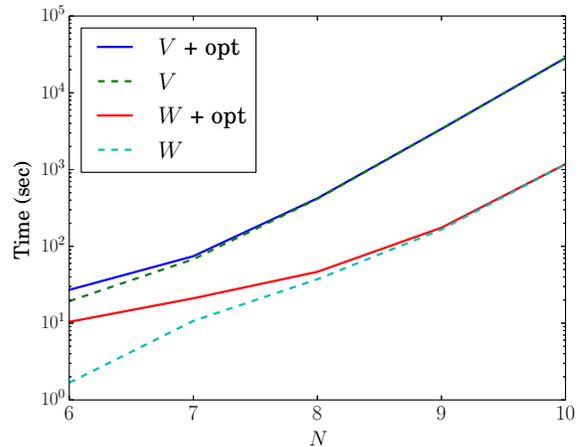}
	}
	\caption{
		Calculation times (in seconds) for obtaining $\mathcal{F}$ and $\mathcal{I}$ for 100 random mixed states. The dashed curves show the elapsed times only for constructing the matrix $V$ or $W$, whereas the solid curves show the results including the elapsed time to get a optimized value. The upper two curves are the results for $\mathcal{F}$ which are steeper than the lower curves which are the results for $\mathcal{I}$.
	}
	\label{fig:time}
\end{figure}

To support our argument, we randomly generated $100$ mixed states with $N=4,6,8,10$, and counted the times taken to calculate $\mathcal{I}$ and $\mathcal{F}$. We tested our code using the single core of the system with Xeon E5-2620 CPU and 64GB memory and averaged ten trials. To get an optimized value, we parametrized $\pmb{\alpha}$ using $2N$ variables and provided $200$ initial random points. For each point of $\pmb{\alpha}$, we used the Broyden-Fletcher-Goldfarb-Shanno method~\cite{fletcher87} to find the local maximum value of the function which is implemented in the GNU Scientific Library~\cite{gsl}. The results in Fig.~\ref{fig:time} directly show that the time cost for $\mathcal{F}$ is much bigger and grows faster than that of $\mathcal{I}$. The results also show that the overhead time for the optimization over the total calculation time decreases as $N$ increases.

We also note that the optimization problem can be readily simplified if a density matrix $\rho$ has the permutation symmetry. In this case, the matrix $V$ (or $W$) has the form $V_{ia,jb}(W_{ia,jb})=D_{ab}$ if $i = j$ and $V_{ia,jb}(W_{ia,jb}) = O_{ab}$ if $i \neq j$ where $D$ and $O$ are $3\times3$ matrices. The optimized value of the measure is given by $N\lambda_1$ where $\lambda_1$ is the maximum eigenvalue of $D+(N-1)O$. Therefore, we do not need an explicit optimization process. 
Still, we need to diagonalize the density matrix to get matrices $O$ and $D$ for the $\mathcal{F}$ case. Therefore, in any cases, we can calculate $\mathcal{I}$ much faster than $\mathcal{F}$.

\section{ Calculation of $\mathcal{I}$ for the ground state of Ising model } \label{sec:transverseising}

The Hamiltonian of the Ising model is given by
\begin{equation}
H_{\tn{Ising}}(\lambda)=-\sum_{j=1}^{N} \left( \lambda \sigma_x^{(j)}\sigma_x^{(j+1)} + \sigma_z^{(j)}\right),
\label{eq:Ising-Hamiltonian}
\end{equation}
where $\sigma_i^{(j)}$ is the $i$-component Pauli operator on the  $j$th site, $\lambda$ is the interaction strength, and  a periodic condition ${\boldsymbol \sigma}^{(N+1)}\equiv{\boldsymbol \sigma}^{(1)}$ is given.
We here briefly review an analytic method to obtain the ground state of this model \cite{Vidal2003} and calculate $\mathcal{I}$ for it.
The Hamiltonian in Eq.~(\ref{eq:Ising-Hamiltonian}) can be diagonalized using 
%well known operator transformations.
%Suppose 
two Majorana operators for each site $l$ of the $N$ spins,
\begin{align}
c_{2l}\equiv \left(\prod_{m=0}^{l-1}\sigma_m^z\right)\sigma_l^x \quad \text{and} \quad c_{2l+1}\equiv \left(\prod_{m=0}^{l-1}\sigma_m^z\right)\sigma_l^y,
\end{align}
where $c_m$ is a Hermitian operator and satisfies the anti-commutation relation, $\{c_m,c_n\}=2\delta_{mn}$.
It is known that the expectation values $\langle c_m c_n \rangle=\delta_{mn} + i \Gamma_{mn}$ completely characterize the block of $L$ spin-$1/2$ particles of the ground state \cite{Vidal2003}, where the matrix $\Gamma$ is has the form,
\begin{align}
&\Gamma=\left[
\begin{array}{cccc}
\Pi_0 & \Pi_1& \cdots& \Pi_{L-1}\\
\Pi_{-1} & \Pi_0 &  & \vdots\\
\vdots & & \ddots& \vdots\\
\Pi_{1-L} & \cdots& \cdots& \Pi_0\\
\end{array}
\right],
\nonumber\\
&\Pi_l = \left[
\begin{array}{cc}
0 & g_l \\
-g_{-l} & 0 
\end{array}
\right],
\end{align}
where $g_l$ is given by 
\begin{equation}
g_l=\frac{1}{2\pi} \int_0^{2\pi} d\phi \ e^{-il\phi} \frac{\lambda e^{-i \phi} -1 }{\abs{\lambda e^{-i \phi} -1}}.
\end{equation}
Since $\Gamma$ is a skew-symmetric matrix, we can find an orthogonal matrix $V\in \tn{SO}(2L)$ that block-diagonalizes $\Gamma$ into
\begin{equation}
\tilde{\Gamma}=V\Gamma V^T = \oplus_{m=0}^{L-1} \nu_m \left[ 
\begin{array}{cc}
0 & 1 \\
-1 & 0
\end{array}
\right].
\end{equation}
The set of $2L$ Majorana operators $d_m = \sum_{n=0}^{2L-1} V_{mn}c_n$ has a block-diagonal correlation matrix $\langle d_m d_n \rangle = \delta_{mn}+i \tilde{\Gamma}_{mn}$.
Now, $L$ fermionic operators $b_l\equiv (d_{2l}+i d_{2l+1})/2$ obeying $\{b_m,b_n\}=0$ and $\{ b_m^\dagger, b_n\}=\delta_{mn}$ have expectation values
\begin{equation}
\langle b_m\rangle = 0, \qquad \langle b_m b_n \rangle = 0, \qquad \langle b_m^\dagger b_n \rangle = \delta_{mn} \frac{1+\nu_m}{2}.
\label{eq:b-component}
\end{equation}
This indicates that the mixed states of the block of length $L$ can be described as a product sate in $b_m$ basis as  
\begin{equation}
\rho_L=\bigotimes_{m=0}^{L-1} \rho_m.
\label{eq:rhoL}
\end{equation}
We re-express Eq.~(\ref{eq:rhoL}) in the Pauli operator basis.
From Eq.~(\ref{eq:b-component}), one can find
\begin{equation}
\rho_m = \frac{1-\nu_m}{2} b_m b_m^\dagger + \frac{1+\nu_m}{2}b_m^\dagger b_m,
\end{equation}
and we can expand $b_m$ as
\begin{align}
b_m &= \frac{1}{2}(d_{2l}+i d_{2l+1}) = \frac{1}{2} \sum_{n=0}^{2L-1} \left( V_{2l,n}c_n + iV_{2l+1,n} c_n \right) \nonumber \\
&=\frac{1}{2}\Big[ \sum_{k=0}^{L-1} \left(V_{2l,2k}+i V_{2l+1,2k}\right) c_{2k} \nonumber \\
& \qquad \qquad + \sum_{k=0}^{L-1} \left(V_{2l,2k+1}+i V_{2l+1,2k+1}\right) c_{2k+1}\Big] \nonumber \\ 
&=\frac{1}{2}\Bigg[ \sum_{k=0}^{L-1} \left(V_{2l,2k}+i V_{2l+1,2k}\right) \left(\prod_{m=0}^{k-1}\sigma_m^z\right)\sigma_k^x \nonumber \\
& \qquad + \sum_{k=0}^{L-1} \left(V_{2l,2k+1}+i V_{2l+1,2k+1}\right) \left(\prod_{m=0}^{k-1}\sigma_m^z\right)\sigma_k^y \Bigg].
\end{align}
After obtaining the density matrix $\rho_L$, we can numerically calculate the values of $\mathcal{I}(\rho_L)$ and $\mathcal{F}(\rho_L)$ using the method in the previous section (Appendix~\ref{sec:time_compare}).

\end{document}